# Design and Measurement of Dipole Magnets for CSNS 1.6GeV RCS*

Li Qing(李青)[1, 2, 3; 1)], Sun Xian-Jing(孙献静)[1], Kang Wen(康文)[1, 2, 3], Deng Chang-dong (邓昌东)[1, 2, 3], Chen Wan (陈宛)[1]

1. Institute of High Energy Physics, Chinese Academy of Sciences, Beijing 100049, China
2. Dongguan Neutron Science center, Dongguan 523803, China
3. Dongguan Key Lab. Of High precision Magnetic Field Measurement, Dongguan 523803, China

**Abstract:** The Rapid Cycling Synchrotron (RCS) in Chinese Spallation Neutron Source (CSNS) accelerates proton beam from 80Mev to 1.6GeV at a repetition rate of 25Hz. All dipole magnets of RCS are operated at AC with biased DC. Aiming at the properties of these dipole magnets, we take some methods to improve magnetic field quality in the good region and reduce eddy currents in the iron core . In this paper, we would present the process of the magnet design and temperature rise calculation. At the same time, the field measurement results and temperature test of the prototype magnet are also described and discussed.

## 1. Introduction

The Rapid Cycling Synchrotron (RCS) is one part of Chinese Spallation Neutron Source (CSNS). It accelerates the proton beam from 80MeV to 1.6GeV. The RCS has a fourfold symmetric lattice. One fourth period consists 6 dipole and 12 quadrupole magnets as shown in Fig.1 [1-2]. Dipoles and quadrupoles are excited by AC with biased DC and the repetition rate is 25Hz. All 24 dipole magnets are excited by one resonant power supply. The final new physical parameters are showed in table 1. In order to study the properties of AC magnets, a dipole prototype had been designed, fabricated and measured before batch production.

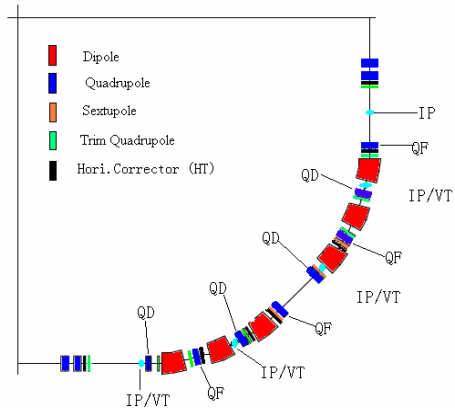

Fig.1 One fourth period of RCS

Table 1: Final new physical parameters of dipole magnet

*Supported by the National Natural Science Foundation of China（11305195）and China Spallation Neutron Source Project

1)   E-mail: liqing@ihep.ac.cn

| Parameter | Value |
| --- | --- |
| Number of magnets | 24 |
| Average radius | 8.021m |
| Gap | 160mm |
| B(0) | 0.1645T (injection) |
| | 0.9807T ( extraction) |
| Magnetic length | 2.1m |
| Excitation frequency | 25Hz |
| Good field size | 212.0mm×134.0mm (injection) |
| | 185.0mm×110.0mm ( extraction) |
| Field quality($\Delta(BL)/(B_0L)$ ) | $<\pm 8\times 10^{-4}$ |
| $\Delta(B_6L)/(B_0L)$ | $<\pm 5\times 10^{-4}$ |
| $\Delta(B_{10}L)/(B_0L)$ | $<\pm 5\times 10^{-4}$ |
| $\Delta(B_{14}L)/(B_0L)$ | $<\pm 6\times 10^{-4}$ |
| Non-linearity of the magnetic field | <1.5% |
| Integral field difference | $<\pm 0.1\%$ |

## 2. Design of the dipole magnet

From table1, the dipole magnets are characterized as follows:
1) Large gap and rapid changing excitation frequency. A large gape which is determined by the strong space-charge effect of proton beam, causes a large leakage field at the end of the magnet. Eddy currents will be induced by leakage field with rapid changing excitation frequency, and then caused high temperature in core end of the dipole magnet [3-4].
2) Large difference between injection and extraction field. It is hard to unify the magnetic field quality in low field and high field.

The whole iron core of the dipole magnet is H-type curved yoke, which is divided into upper core and lower core, shown in Fig.2. Cross section design is optimized by the OPERA/2D. The height and width are 1160mm and 1430mm. The width of the pole is 476mm. At the pole corner, a bump shape is optimized to improve the field quality. Fig.3 shows the 1/4 cross section and the field distribution on the middle line.

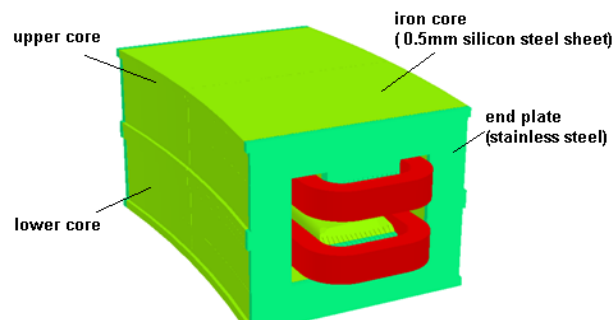

Fig.2 H-type curved yoke

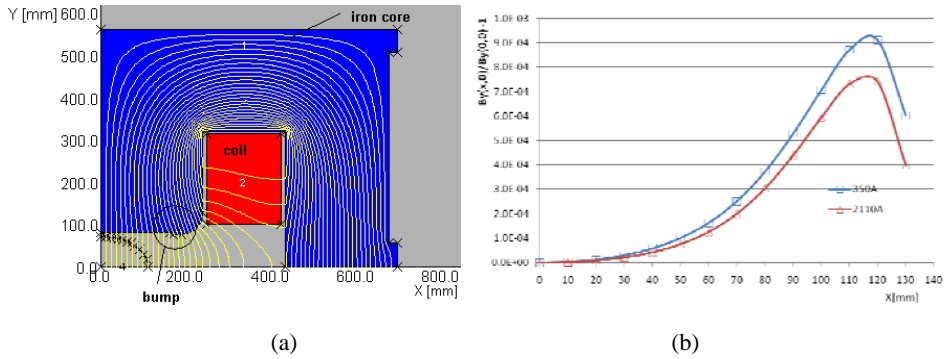

(a)                 (b)

Fig.3 (a) 1/4 cross section    (b) The field distribution on the middle line

After determining the cross section, the OPERA/ 3D model is built by extruding the cross section in parallel, which is shown in Fig.2. The arc radius of the curved yoke is 8021mm. The iron core is stacked with 0.5mm silicon steel sheets and 40mm thick stainless steel plates at the ends of iron core. At the end parts of pole, a Rogowski curve is applied. There are two benefits of doing this:

1) Reducing the saturation of ends.

2) Reducing the magnetic field component $B_z$ which perpendicular to the end plate. The eddy current that causes heat up is induced mainly by the magnetic field component $B_z$ [5].

Considering machine the curve easily, we choose three lines to approximate the curve, shown in Fig4. In order to meet the requirement of integral field uniformity, the end chamfering is necessary at the end pole. We use the harmonic chamfering method to optimize the end chamfering shape. This method can counteract each multipole error through the change of pole profile and improves the field quality radically [6]. The chamfering shape is shown in Fig.5.

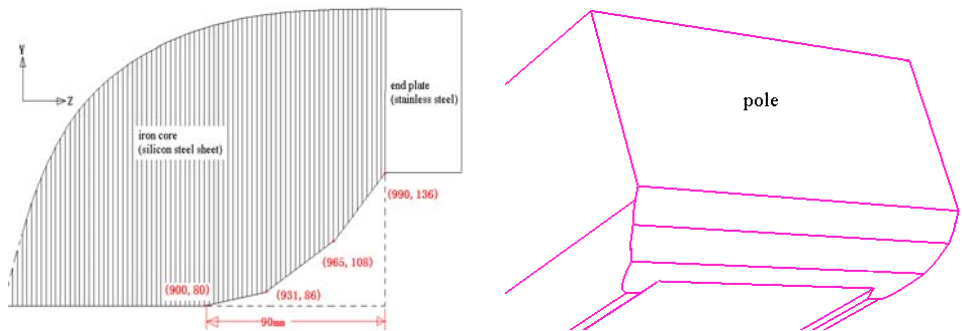

Fig.4 Rogowski three lines

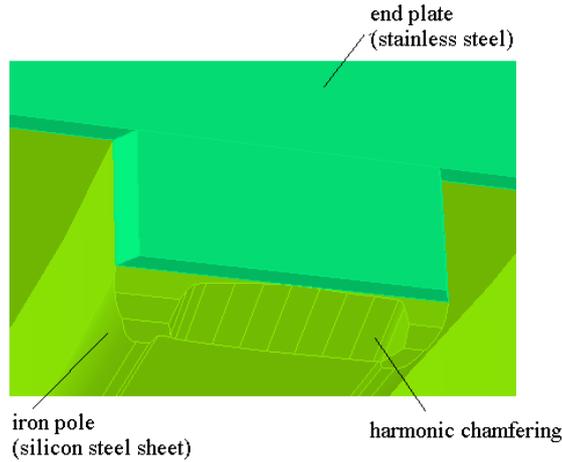

Fig.5 The chamfering shape on the pole

The non-linearity of center magnetic field B(0) and center integral field BL(0) are shown in Fig.6. The non-linearity of BL(0) is 1.2% for the magnet field. In good field region, the uniformities are less than $\pm 5\times 10^{-4}$. The high order harmonic fields were less than $\pm 3.2\times 10^{-4}$. Detailed distribution is showed in Fig.7.

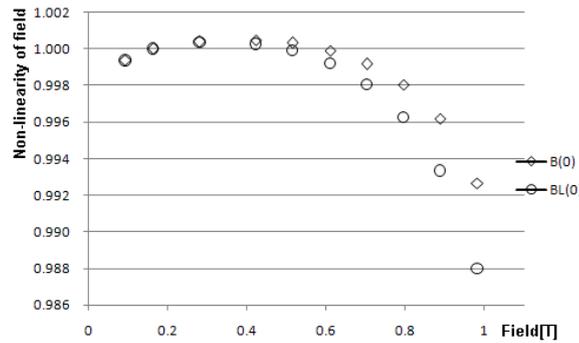

Fig.6 The non-linearity of B(0) and BL(0)

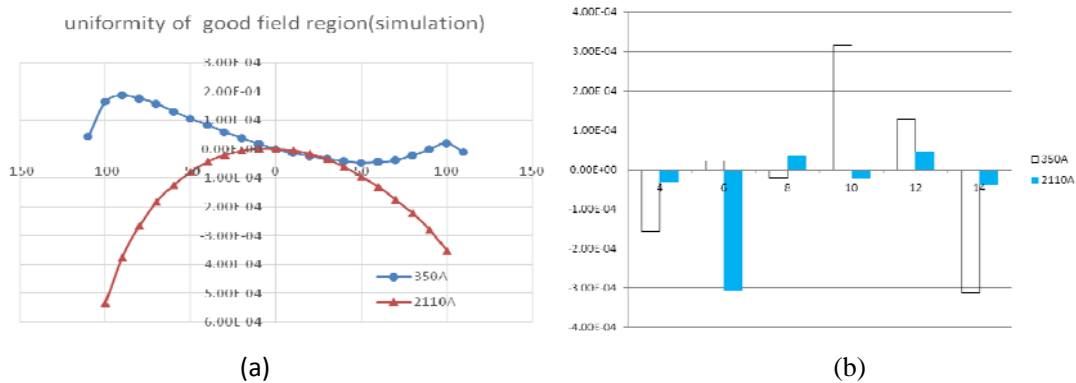

(a)          (b)

Fig.7 (a) The uniformity of good field. (b) The high order harmonic fields.

Not only does the eddy currents exist in stainless steel end plates, but also exists in the iron core, especially at the end part [4].The heat-up caused by eddy currents should be calculated carefully. Otherwise, the temperature will rise rapidly to break the epoxy. The simple and effective method is to arrange a series of slits which is perpendicular to the end. The optimized goal is the maximum temperature less than 130℃. ELEKTRA-TR of OPERA/3D is used to calculate the flux density and eddy current losses. Then in the TEMPO-ST model of OPERA/3D, the loss data are put into to obtain the temperature distribution [7-8]. The excitation current is shown in Fig.8.

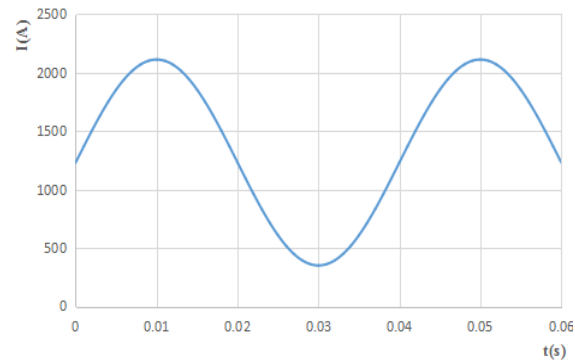

Fig.8 The excitation current DC biased AC

Fig.9 shows the eddy current flow (@0.05s) and the final temperature distribution before slitting. From Fig.9 we can see that the maximum temperature is 202℃ and located in the middle of stainless steel end plate. The comb type slits were arranged in the place of large eddy current loop. Fig.10 shows the eddy current flow (@0.05s) and temperature distribution. By optimizing design, the maximum temperature is less than110℃ and located at the inner arc corner of magnet pole end.

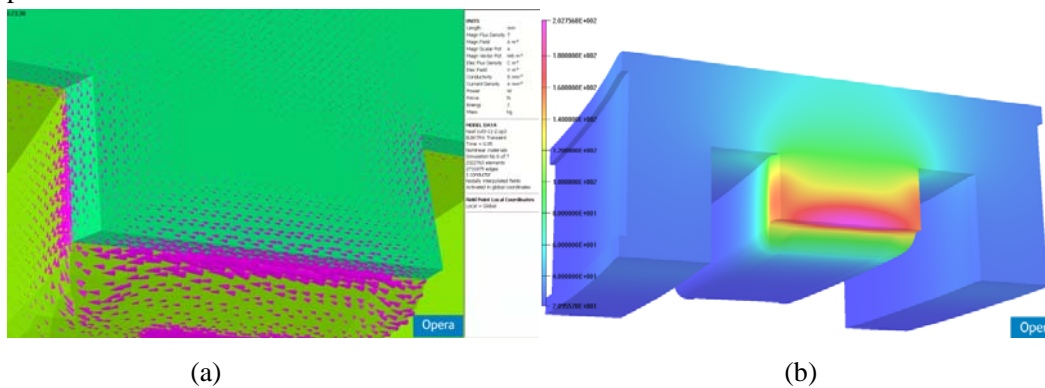

(a)                  (b)

Fig.9 (a)Eddy current flow(@0.05s). (b) Temperature distribution before slitting.

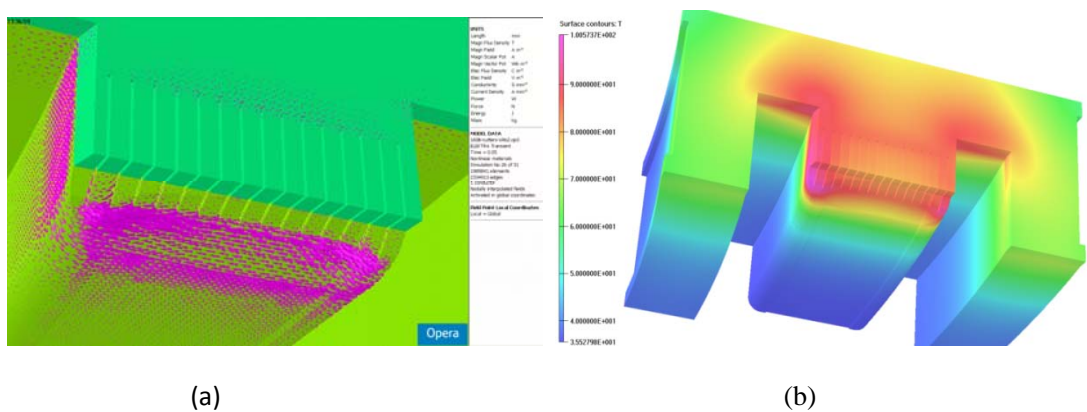

(a)                  (b)

Fig.10 (a) Eddy current flow(@0.05s). (b) Temperature distribution after slitting.

In the high repetition magnet, the aluminum stranded conductors have been developed to wind the coils[3]. The square cross section size is 34mm×34mm with 72 filaments of 3mm diameter inside[2]. The cross section of conductor and coil are shown in Fig.11. This kind of conductor can reduce the eddy current in the coils and have lower inductance.

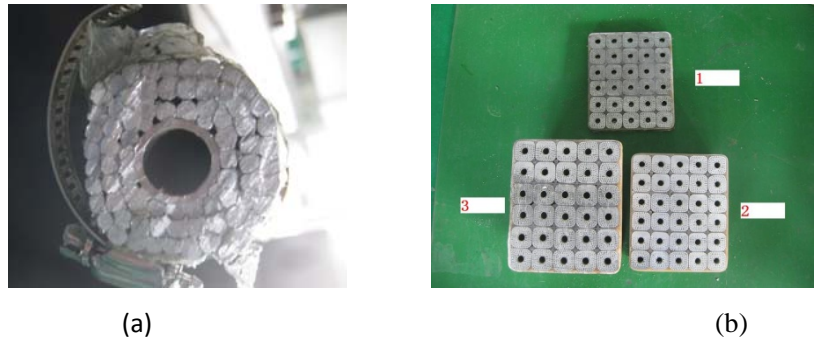

(a)                                                         (b)

Fig.11 (a) Cross section of conductor. (b) Cross section of coil.

## 3. Measurement of prototype dipole magnet

### 3.1 Magnetic field measurement

The prototype dipole magnet with new parameters was fabricated and tested carefully. The magnetic field was measured by a long search coil system. This system contains a Hall probe, a long arc test coil with 250 turns for DC, a long straight test coil with for AC(biased DC ) and some other electronic equipments. The DC field distribution along center arc was tested by Hall probe. The uniformity of integral field was measured by long arc test coil. The uniformity and high order harmonic fields are showed in Fig12. The results are all less than $\pm 5 \times 10^{-4}$. The non-linearity of DC field is 0.64% which is shown in Fig.13. The non-linearity of AC(biased DC ) was measured by flipping the long straight coil, the non-linearity is 1.88% which is shown in Fig.14. All these results meet the requirements.

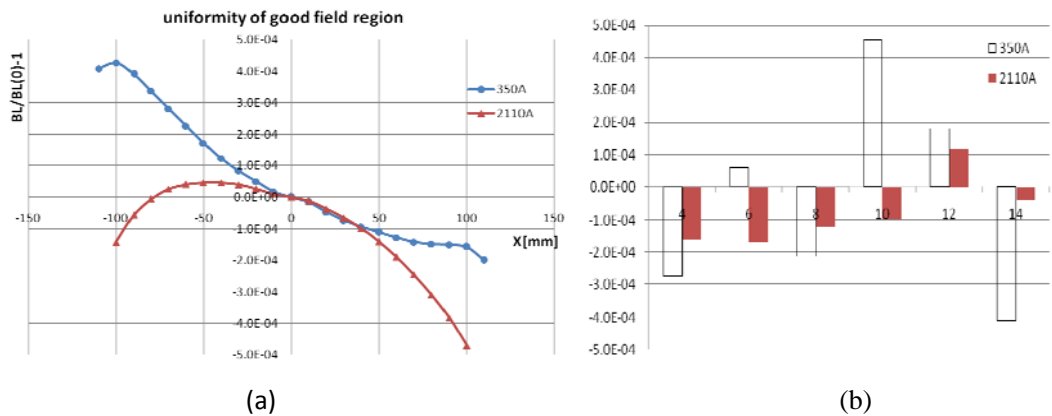

(a)                                                         (b)

Fig.12 (a)The uniformity of good field (measurement). (b) The high order harmonic fields(measurement).

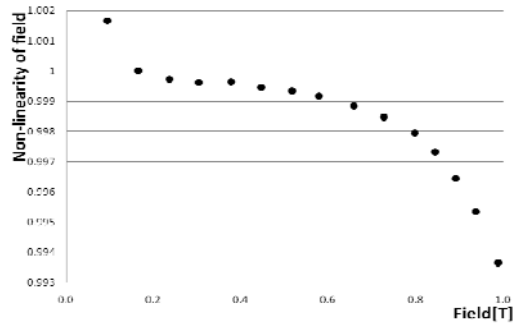

Fig.13 The non-linearity of DC field (measurement)

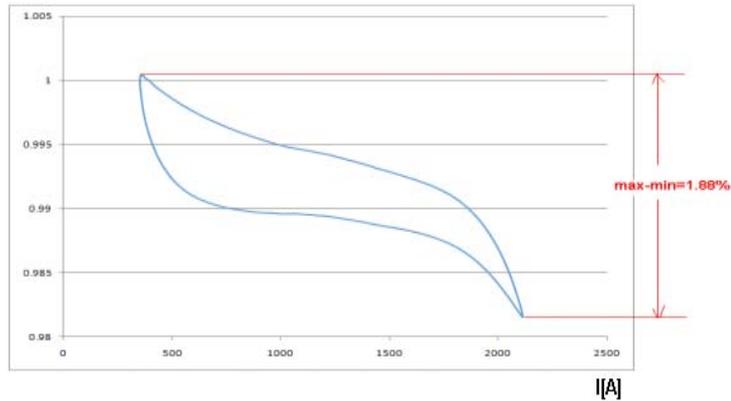

Fig.14 The non-linearity of AC(biased DC ) field (measurement)

**3.2 Temperature measurement**

The prototype magnet is excited by DC1230A±AC880A, and then we use the thermal imager to measure the surface temperature. The whole measurement process lasted 93 hours(from 21.4h to 24.4h, we unloaded the current for some reasons ). The highest temperature appeared on the inner arc corner of the pole ends, shows in Fig.15. There are 4 inner arc corners and their temperature changes were recorded in Fig.16. About after 40 hours, the temperature reached a balance and no longer rising. The final maximum temperature was less than 130℃.

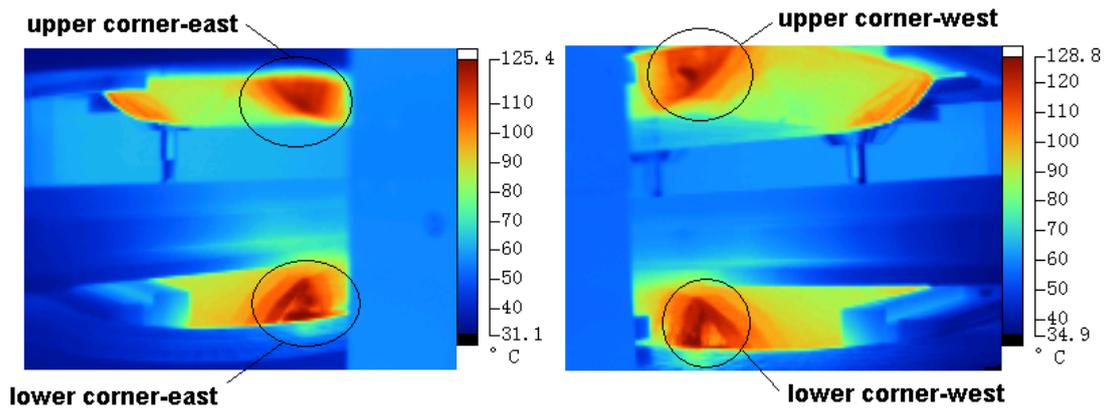

Fig.15 4 inner arc corners temperature with thermal imager

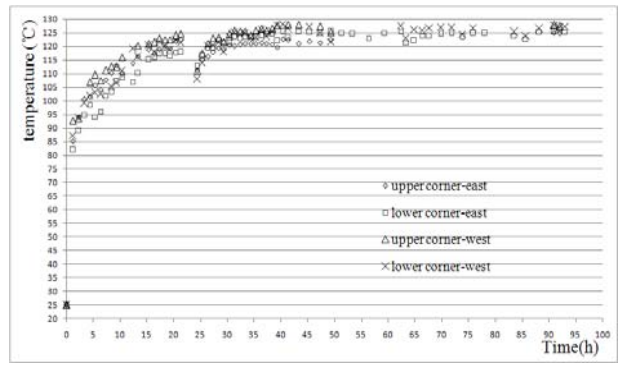

Fig.16 4 inner arc corners temperature changes

## 4. Conclusion

In magnet design, the harmonic chamfering method is applied to improve the field quality radically. Through the electromagnetic field simulation and the thermal analysis, the optimized slits arrange is adopted and the eddy current in iron core ends are reduced sharply. The prototype magnet has been test successfully. The highest temperature has been controlled lower than 130℃, and it is in the allowable region of temperature. From the measured results of magnetic fields, the field quality met the requirement well. The uniformity and the high order harmonic field in the good field region were less than $\pm 5 \times 10^{-4}$.